\documentclass[runningheads,natbib]{svjour2}
\bibpunct{[}{]}{;}{n}{}{,} 
\smartqed  
\usepackage{graphicx}
\journalname{Space Science Reviews}
\begin{document}

\title{Earthshine observation of vegetation and implication for life detection on other planets
}
\subtitle{A review of 2001 - 2006 works}

\titlerunning{Vegetation as a global biomarker}        

\author{Luc Arnold     
}


\institute{
              Observatoire de Haute Provence CNRS 04870 Saint-Michel-l'Observatoire, France \\
              Tel.: +33 4 92 70 64 00\\
              Fax: +33 4 92 76 62 95\\
              \email{Luc.Arnold @ oamp.fr}           
}

\date{Received: 23 Feb 2007 - Revised: 26 Jun 2007}

\maketitle

\begin{abstract}

The detection of exolife is one of the goals of very ambitious future space missions that aim to take direct images of Earth-like planets. While associations of simple molecules present in the planet's atmosphere ($O_2$, $O_3$, $CO_2$ etc.) have been identified as possible global biomarkers, we review here the detectability of a signature of life from the planet's surface, i.e. the green vegetation. The vegetation reflectance has indeed a specific spectrum, with a sharp edge around 700 nm, known as the "Vegetation Red Edge" (VRE). Moreover vegetation covers a large surface of emerged lands, from tropical evergreen forest to shrub tundra. Thus considering it as a potential global biomarker is relevant. 

Earthshine allows to observe the Earth as a distant planet, i.e. without spatial resolution. Since 2001, Earthshine observations have been used by several authors to test and quantify the detectability of the VRE in the Earth spectrum. The vegetation spectral signature is detected as a small 'positive shift' of a few percents above the continuum, starting at 700 nm. This signature appears in most spectra, and its strength is correlated with the Earth's phase (visible land versus visible ocean). The observations show that detecting the VRE on Earth requires a photometric relative accuracy of 1\% or better. Detecting something equivalent on an Earth-like planet will therefore remain challenging, moreover considering the possibility of mineral artifacts and the question of 'red edge' universality in the Universe.

\end{abstract}

\newpage

\section{Introduction}
\label{intro}

\textit{Shall we be able to detect life on an unresolved Earth-like extrasolar planet ?} 
Future space missions like Darwin or TPF will provide us with the first images and low-resolution spectra of these planets, and the question of the presence or not of -ideally- an unambiguous biosignature or -more realistically- a set of possible biogenic spectral features in these data will undoubtedly feed an animate debate. 

Let us consider a unresolved extrasolar Earth-like planet imaged by a space-based high-contrast instrument, basically a telescope equipped with a coronagraph that blocks the stellar light by masking the star to allow the observation of the very faint planet near its parent star \footnote{An earth-like planet is $\approx 10^{10}$ fainter at visible wavelengths than its parent star.}. The spectrum of the light reflected by the planet, when normalized to the parent star spectrum, gives the planet reflectance spectrum revealing its atmospheric and ground colour, if the latter is visible through a partially transparent atmosphere. Since the planet will remain unresolved (at least with the missions mentioned above), its spectrum will be spatially integrated (i.e. disk-averaged) for the observed orbital phase of the planet. 

\begin{figure}
\centering
  \includegraphics{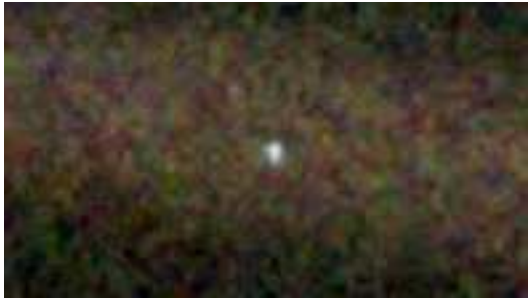}
\caption{Voyager-1 took this picture of Earth in February 1990 while it was traveling well beyond the orbit of Neptune. Voyager did not take a spectrum of this spatially unresolved view of its mother planet, but this picture illustrates how could look like an Earth-like extrasolar planet imaged in the visible domain by a future space observatory - a pale blue dot. Photo from http://photojournal.jpl.nasa.gov/catalog/PIA02228. Courtesy NASA/JPL-Caltech.}
\label{voyager}
\end{figure}

\begin{figure}
\centering
  \includegraphics[width=1.\textwidth, height=7cm]{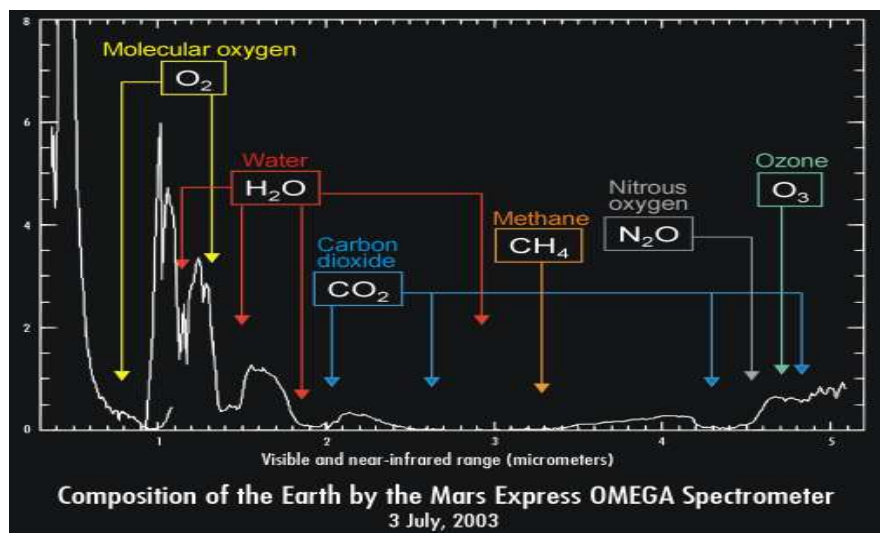}
\caption{Mars Express recorded the Earth spectrum with its OMEGA instrument in July 2003 while it was traveling to Mars. This picture illustrates how could look like an Earth-like extrasolar planet spectrum recorded with a high signal to noise ratio (figure adapted from http://mars.jpl.nasa.gov/express/newsroom/pressreleases/20030717a.html).}
\label{omega}
\end{figure}

\textit{How would the spectrum of an unresolved Earth-like planet look like?} A way to answer this question is to consider how would look like the spectrum of our Earth
if it would be observed from a very large distance, typically several parsecs. This can be done from a space probe traveling deep into the Solar System and looking back the Earth, as Voyager-1 did in 1990 (Fig.~\ref{voyager}) or Mars Express in 2003 (Fig.~\ref{omega}). 
Note also that an integrated Earth spectrum for a given phase of the planet and also for a given position of an observer far above the earth  can also be done - in principle at least - by integrating spatially-resolved spectra from low-orbit satellites. 

\begin{figure}
\centering
  \includegraphics[width=0.5\textwidth, height=6.5cm]{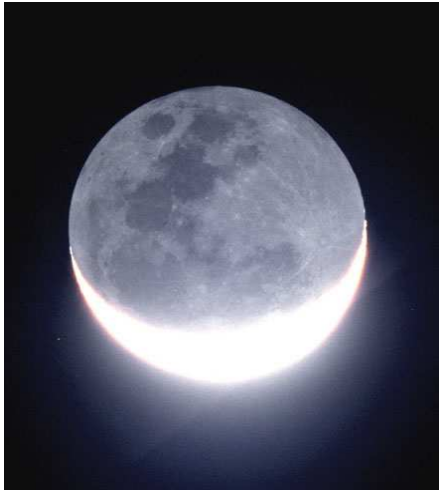}
\caption{The Moon, in a morning of September 1999, displaying its crescent, i.e. the sunlit part of the Moon (here overexposed), and also a bright Earthshine over the rest of the Moon disk, i.e. the non-sunlit Moon illuminated by a gibbous Earth. It seems that Leonardo Da Vinci has been the first who clearly understood the origin of the phenomenon of Earthshine when studying the geometrical relationship between the Earth, Moon and Sun \cite{welther99}. The spectroscopy of the Earthshine directly gives the disk-averaged spectrum of the Earth as seen from the Moon \textit{(photo Luc Arnold)}.}
\label{moon}
\end{figure}

An alternative method to obtain the Earth averaged spectrum consists of taking a spectrum of the Moon Earthshine, i.e. Earth light backscattered by the non-sunlit Moon (Fig.~\ref{moon}). A spectrum of the Moon Earthshine directly gives the disk-averaged spectrum of the Earth at the phase seen from the Moon (since the Moon surface roughness "washes out" any spatial information on the Earth's colour).

\textit{What shall we look for in this spectrum ?}
We shall first look for sets of molecules in the planet atmosphere (like oxygen and ozone) which may be biologic products or by-products. We shall also look for ground colours characteristic of biological complex molecules (like pigments in vegetation). Said in a more general manner, we shall look for missing photons used in a photosynthetic process occurring on the planet.  
The visible and near infra-red Earthshine spectra published to date clearly show the atmospheric signatures and, at least, tentative signs of ground vegetation which thus appears as an interesting potential global biomarker \cite{arnold02}, \cite{woolf02}, \cite{seager05}, \cite{montanes05},\\ \cite{montanes06}, \cite{hamdani06}. The question of possible artifacts is of course of prime importance (see Sect. \ref{implication}).

\begin{figure}
\centering
  \includegraphics[width=0.8\textwidth, height=5cm]{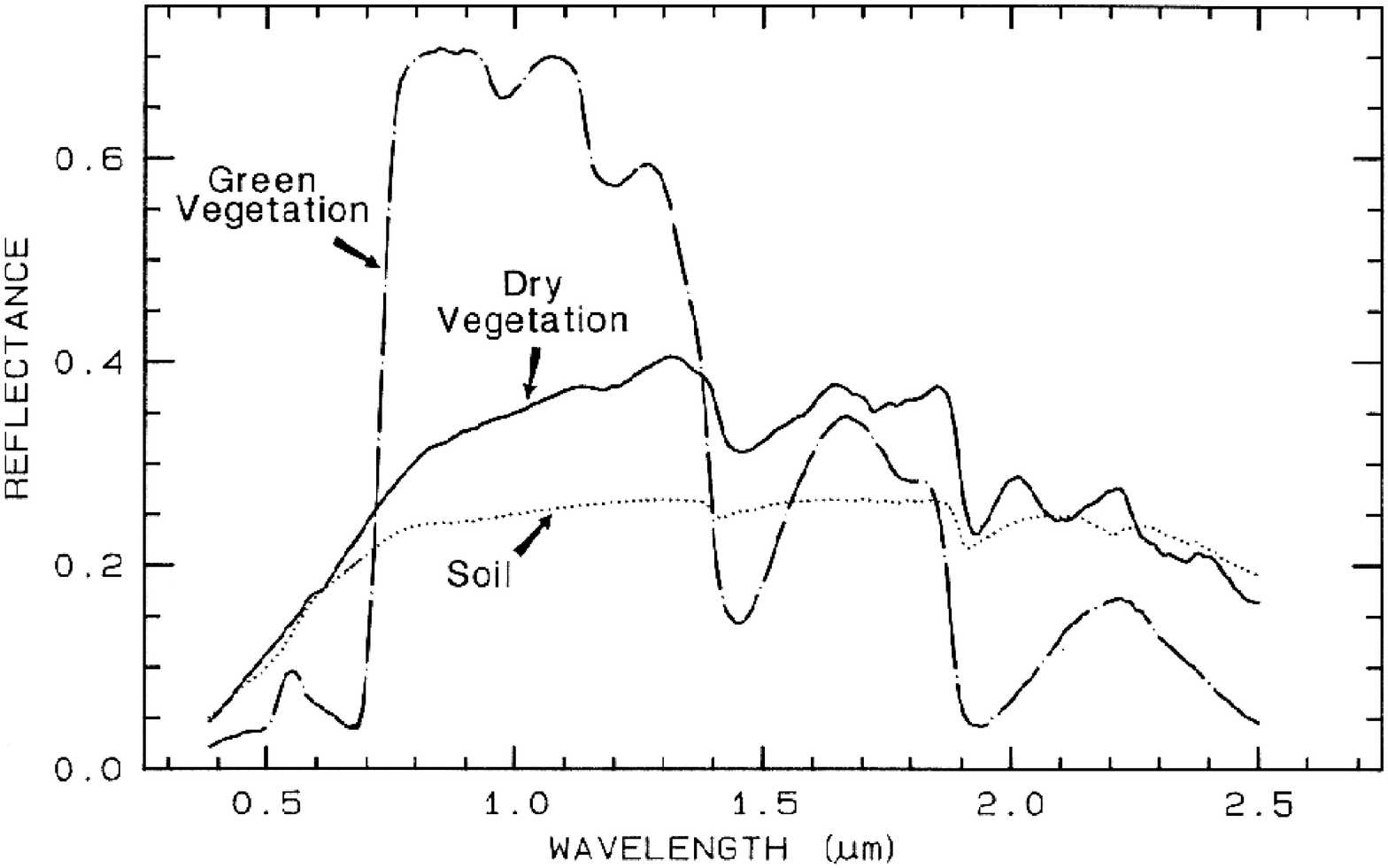}
\caption{Reflectance spectra of photosynthetic (green) vegetation, non-photosynthetic (dry) and soil \cite{clark99}.
   The so-called vegetation red edge (VRE) is the green vegetation reflectance strong increase from $\approx5\%$ at $670\ \rm nm$
   to $\approx70\%$ at $800\ \rm nm$.}
\label{vegetation_spectrum}
\end{figure}

Vegetation indeed has a high reflectivity in the near-IR, higher than in the visible by a factor of $\approx5$ (\cite{clark99}, Fig.~\ref{vegetation_spectrum}). This produces a sharp edge around $\approx 700\ \rm nm$, the so-called Vegetation Red Edge (VRE). An Earth disk-averaged reflectance spectrum is thus expected to rise by a significant fraction around this wavelength if vegetation is in view from the Moon when the Earthshine is observed. 

The two next sections of this paper present the basics about Earthshine spectroscopy and the results about VRE measurements collected between 2001 and 2005.
Sections~\ref{learn} and \ref{implication} discuss these results, the perspective and implication for life detection.

\section{Basics on Earthshine spectroscopy and vegetation red edge signal}
\label{review_res}
It seems that Leonardo Da Vinci has been the first who clearly understood the origin of the phenomenon of Earthshine when studying the geometrical relationship between the Earth, Moon and Sun \cite{welther99}. The potential of Moon's Earthshine in providing global data on Earth has been identified during the 19$^{th}$ century \cite{flammarion}, maybe even earlier.  Arcichovsky suggested in 1912 to look for chlorophyll absorption in the Earthshine spectrum, to calibrate chlorophyll in the spectrum of other planets \cite{arci1912}. This approach re-emerged again within the context of the preparation of Darwin and TPF missions, when J. Schneider from Paris-Meudon Observatory suggested new observations done at ESO in 1999 and at OHP in 2001 \cite{arnold02}. Simultaneous observations have been done by Woolf et al. (2002).

\textit{How is basically Earth spectrum extracted from the Earthshine ?}
Let us call the Sun spectrum as seen from outside the Earth atmosphere $S(\lambda)$,
Earth atmosphere transmittance $AT(\lambda)$, Moonlight
$MS(\lambda)$, Earthshine $ES(\lambda)$, \\ 
Moon reflectance
$MR(\lambda)$, and Earth reflectance $ER(\lambda)$. We have
\begin{eqnarray}
MS(\lambda) = S(\lambda)\times MR(\lambda) \times AT(\lambda) \times g_1,
\label{MS}
\end{eqnarray}
\begin{eqnarray}
ES(\lambda) = S(\lambda)\times ER(\lambda)\times MR(\lambda) \times
 AT(\lambda) \times g_2.
\label{ES}
\end{eqnarray}
The Earth reflectance is simply given by the ratio Eq.\ref{ES}/Eq.\ref{MS}, i.e.
\begin{eqnarray}
ER(\lambda) = {ES(\lambda) \times g_1 \over MS(\lambda) \times g_2}.
\label{EA}
\end{eqnarray}
Simplifying the ratio by $AT(\lambda)$ means that $ES(\lambda)$ and $MS(\lambda)$ should be ideally recorded simultaneously to avoid significant airmass variation and thus an incorrect Rayleigh scattering measurement. The mean of the two $MS$ spectra bracketing $ES(\lambda)$ is thus used to compute $ER(\lambda)$. The $g_i$ terms are geometric factors related to the geometry of the Sun, Earth and Moon triplet. For simplicity, the terms $g_1$ and $g_2$ are set to 1, equivalent to a spectrum normalization. 
The previous equations assume the sky background has been properly subtracted from the spectra. The data reduction, either for broadband photometry or spectroscopy, is described in \cite{arnold02, qiu03, turnbull06, hamdani06}. The VRE is extracted from $ER(\lambda)$ and defined by the ratio
\begin{equation}
VRE=\frac{r_I-r_R}{r_R} \label{VRE}
\end{equation}
where $r_I$ and $r_R$ are the near-infrared (NIR) and red reflectance integrated over given spectral domains ($\approx 10$ nm width).

\section{Review of results}
\label{review_res}

Table \ref{tab_vre} presents the VRE values collected from the literature. Observations roughly confirm what Schneider (2002) and Des Marais et al. (2002) had inferred from
their previous conjecture, i.e. that vegetation signature is detectable in an integrated (or disk-averaged) Earth spectrum. Observations showed that this signature
is weak and variable as suspected, depending, for example, from the ratio between ocean and land in view from the Moon at the time of the observation, or the cloud cover above vegetated area.  

Seager et al. (2005) present two Earthshine spectra recorded on an evening and morning Moon respectively, thus with an Earth in view from the Moon significantly different
for the two observations. Although their results remain unfortunately only qualitative, the morning spectrum (South America in view) seems to show a weak signal around 700 nm, while the evening spectrum (Pacific ocean in view) remains flat. Hamdani et al. (2006) with spectra recorded from  Chile (ESO NTT) also observe a lower VRE of 1\% when mainly Pacific ocean was in view, and and higher VRE of 3 to 4\% when a significant land area is in view. It is worthwhile to note that Europe-based Earthshine observers always have significant land in view from the Moon. Woolf et al. (2002) announce a VRE of 6\%, which may be overstimated, considering that a large part of ocean was in view when the observation was done. One of the authors \cite{jucks02} said the VRE may probably be closer to 3\%, although apparently never confirmed in a subsequent paper.

Recent results \cite{hamdani06, montanes06} suggest that the results from our very simple model described in Arnold et al. (2002) are overestimated too, as well as the VRE measured from POLDER data \cite{arnold03}, which clearly are biased by desert, as explained later in this paper (Sect. \ref{learn}). It must also be noted that the highest of our measured VRE value of 10\% has an estimated error of $\pm5\%$ \cite{arnold02}, suggesting that our most significant values are rather VRE=4 and 7\%, for which the  measurements have a better accuracy. 

Monta\~ n\' es-Rodriguez et al. (2005) observe the Earthshine the 19th of November 2003 and conclude from a first analysis that no sign of vegetation is visible
in their spectra. But they later re-analyse their spectra \cite{montanes06} with just-released global cloud cover data and conclude that the spectral variations around 700 nm are correlated with cloud-free vegetated area. They obtain VRE values ranging from 2 to 3\% while South and North America are in view.

It is also worthwhile to mention the other features of the Earth spectrum revealed by the Earthshine observations. In addition to the VRE, the red side [600:1000 nm] of the Earth reflectance spectrum shows the presence of $O_2$ and $H_2O$ absorption bands, while the blue side [320:600 nm] clearly shows the Huggins and Chappuis ozone ($O_3$) absorption bands \cite{hamdani06}. The higher reflectance in the blue shows that our planet is blue due to Rayleigh scattering in the atmosphere, as nicely demonstrated by Tikhoff (1914) and Very (1915), and confirmed later with accurate Earthshine broadband photometry by Danjon (1936).

\begin{figure}
\centering
\includegraphics[height=6cm,width=\textwidth]{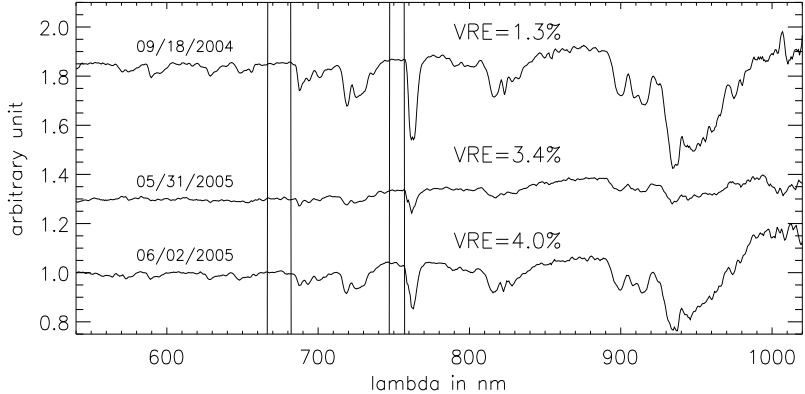}
      \caption{Resulting ER($\lambda$) corrected for $O_3$ Chappuis absorption, Rayleigh and aerosol scattering, thus ready for the VRE measurement. Vertical lines define the two spectral bands used to calculate the VRE. The plots have been shifted vertically for clarity \cite{hamdani06}.}
       \label{vre_hamdani06}
\end{figure}

Clearly ozone absorption from the wide Chappuis band and Rayleigh scattering strongly impact the global shape of the spectrum and needs to be corrected to access the ground 'colour'. Once this is done, the resulting spectrum still contain $O_2$ and $H_20$ bands, but these bands only affect the spectrum locally without compromising the extraction of the VRE (Fig.~\ref{vre_hamdani06}).

\begin{table}[t]
\caption{VRE values from spectroscopy or models. Variations are due to measurements but also to Earth phase (more land or more ocean in view from the Moon at the time of observation - see the papers for details).}
\centering
\label{tab_vre}       
\begin{tabular}{lll}
\hline\noalign{\smallskip}
VRE (\%) & author & method  \\[3pt]
\tableheadseprule\noalign{\smallskip}
5 & Schneider 2000a,b & model\\
$\ge$ 2 & Des Marais et al. 2002 & model\\
4 to 10 & Arnold et al. 2002 & observations \\
7 to 12 & Arnold et al. 2002 & model \\
6 to 11 & Arnold et al. 2003 & POLDER data\\
6 (or 3?) (a)& Woolf et al. 2002 & observations \\
0 to $?$ $(b)$ & Seager et al. 2005 & observations\\
0 & Monta\~ n\' es-Rodriguez et al. 2005 & observations\\
2 to 3 & Monta\~ n\' es-Rodriguez et al. 2006 & observations\\
1 to 4 & Hamdani et al. 2006 & observations\\
$(c)$ & Tinetti et al. 2006 & model\\
$(d)$ & Paillet 2006 & model\\
\noalign{\smallskip}\hline
\end{tabular}
\begin{list}{}{}
\item[$(a)$] Mostly Pacific ocean was in view when this observation was made, therefore the VRE=6\% was maybe overestimated. One of the author \cite{jucks02} said it was probably closer to 3\%.
\item[$(b)$] Not-quantified (weak) uncertain signal when South America in view, but clearly no VRE when mostly ocean in view.
\item[$(c)$] NDVI estimator is used rather than VRE. Conclusion is that vegetation remains detectable on a 24-h averaged Earth observed at dichotomy with a realistic cloud cover. High signal to noise S/N spectra are considered.
\item[$(d)$] Vegetated areas should be 10\% from visible cloud-free surface to be detectable in spectra with signal to noise ratio $S/N \ge10$.
\end{list}
\end{table}

\section{What did we learn from ES observations ?}
\label{learn}

The observations and simulations (Table~\ref{tab_vre}) were all very instructive. If the detection of vegetation in the Earth spectrum is not a surprise,  the VRE remains a small spectral feature, in the 0-to-10\% range above the red continuum (depending on Earth phase, clouds, seasons, position of observer - ocean or land in view, etc). Vegetation has a sharp edge at 700nm but it is easily hidden by clouds (60\% typical cloud cover).

Observers know that Earthshine data reduction remains difficult (although possible with efforts !). Data on Earthshine can have low S/N ratio because they are recorded with the Moon often low above the horizon (high air-mass, low Earthshine fluxes with respect to blue sky background), and on the other hand, the detector can easily be saturated when it records the spectrum of the sunlit Moon crescent. The correction of the pollution of Earthshine by light scattered by the bright Moon crescent (Rayleigh scat.) is one of the key point in the data reduction. Moon colour also varies with phase, moreover Earthshine and crescent are not observed at the same phase angles ! These points are discussed by \cite{hamdani06}.

\begin{figure}
\centering
\includegraphics[height=7.5cm,width=\textwidth]{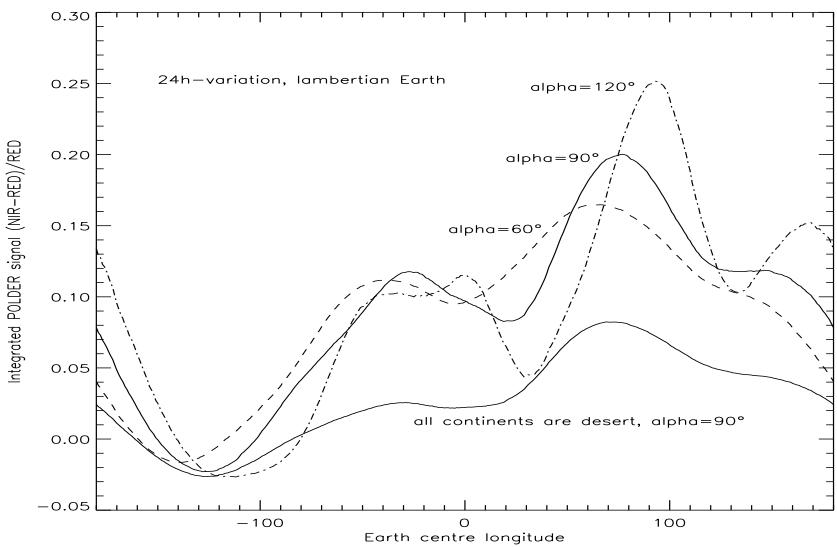}
      \caption{VRE variation over a full Earth rotation. The observer is above latitude $0^{\circ}$ and measures the VRE while the Earth rotates, for
      three different Earth phases (60, 90 and 120$^{\circ}$, respectively gibbous, dichotomy and crescent Earth). The lowest curve simulates a desert Earth seen at phase 90$^{\circ}$: all land pixels have the properties of the Sahara, oceans and clouds are unchanged. The figure shows that the VRE estimator based on 2 bands photometry by POLDER is biased: for the Earth seen at a phase angle of $90^{\circ}$, it shows vegetation if it is $\ge\sim 9\%$, while negative values indicate the presence of ocean.}
       \label{polder}
\end{figure}

\textit{Is broadband photometry, rather than spectroscopy, sufficient to detect vegetation on an unresolved planet ?} To quantify the vegetation signature in the spectrum, at least two estimators can be used. The NDVI (Normalized Difference Vegetation Index) \cite{rouse_et_al74, tucker79}, routinely used for Earth satellite observation, considers the difference, after atmospheric correction, between the reflected fluxes $f$ in broad red and infra-red bands, normalized to the sum of the fluxes in these bands,
\begin{equation}
NDVI =  \frac{f_I-f_R}{f_R+f_I}  \label{NDVI}.
\end{equation}
It can also be written in terms of reflectance, as for the VRE above,
\begin{equation}
NDVI =  \frac{r_I-r_R}{r_R+r_I}.  \label{NDVI}
\end{equation}
But it seems that the NDVI and VRE estimators based on two bands are not sufficient to detect vegetation on an unresolved planet. A large Sahara-like desert can indeed produce a signal similar to a smaller but greener patch on the Earth (see Fig.~\ref{vegetation_spectrum}). Details are given in \cite{arnold03} and Fig.~\ref{polder} shows the VRE variations over 24-h based on POLDER data \cite{deschamps94} for the Earth and an Earth where lands were all attributed to deserts, with clouds and oceans unchanged: The VRE for the latter still indicates the presence of vegetation! Thus only two photometric bands may not avoid a false positive detection on an unresolved exo-Earth, and it is necessary to have a full spectrum to identify the vegetation red edge around 700  nm, with a spectral resolution $\ge\approx50$. A spectrum will allow to distinguish the VRE from a smoother positive slope due to a large desert. The GOME experiment \cite{burrows99} provides spectra well suited to such simulations. It is important to point out that Earth images reconstructed from satellite data are approximation only, although the approximation is probably acceptable for our purpose: Since the data are often recorded for a given solar angle by nadir instruments, it is not possible to take into account effects like shadowing between plants. Nevertheless, at least in principle, a more advanced model can take into account the presence of a hot spot (strong backscattering), or more generally, the profile of the Bidirectional Reflectance Distribution Function (BRDF) for each biome and cloud. Otherwise the Earth is considered as a simple Lambertian diffuser.

\section{Implications for life detection on extrasolar planets: Perspective and open questions}
\label{implication}

The Table~\ref{tab_vre} shows that the VRE is a small feature (a few \% above the continuum) and therefore will require high S/N ratio to be detected.
To be detected by a space-based observatory, the exposure times should of the order of 100 hours to reach S/N=100 with a spectral resolution of 25 for an Earth at 10 pc \cite{arnold02}.

\textit{Any chance to observe a higher VRE on an Earth-like planet than on Earth ?} 
It is quite interesting to note that leaf reflectance of plants increases with leaves thickness \cite{slaton01}. Desert plants with fleshy green stems, often without leaf (in the sense of the common conception of a leaf), generally reflect substantially more radiation than do other plants \cite{gates65}, up to a factor of 2 at 750 nm. The cloud cover over deserts being smaller than the mean cover, desert plants should in principle contribute significantly to the VRE. A planet without much water (a small ocean of albedo $\approx0.1$) and few clouds (albedo $\approx0.6$) would have an albedo dominated by the desert ($\approx0.3$). The planet albedo would thus be roughly the same than Earth's albedo, i.e. $\approx0.3$. Therefore, and paradoxically, such a dry planet, where the majority of plants would have evolved toward a wide variety of cold and warm desert plants might display a stronger VRE than the Earth. Quantitatively speaking and based on GOME spectra of Earth biomes, the VRE could reach $\approx35\%$ for a 50\% desert -plants-covered super-continent\footnote{I mean no ocean in view.}; it would be $\approx15\%$ for a toundra-like super-continent. These comfortable numbers -on the detectability point of view- probably are very optimistic and should be kept in mind as upper limits. Next paragraph may temperate reader's anthusiasm (including mine).

\begin{figure}
\centering
\includegraphics[height=5cm]{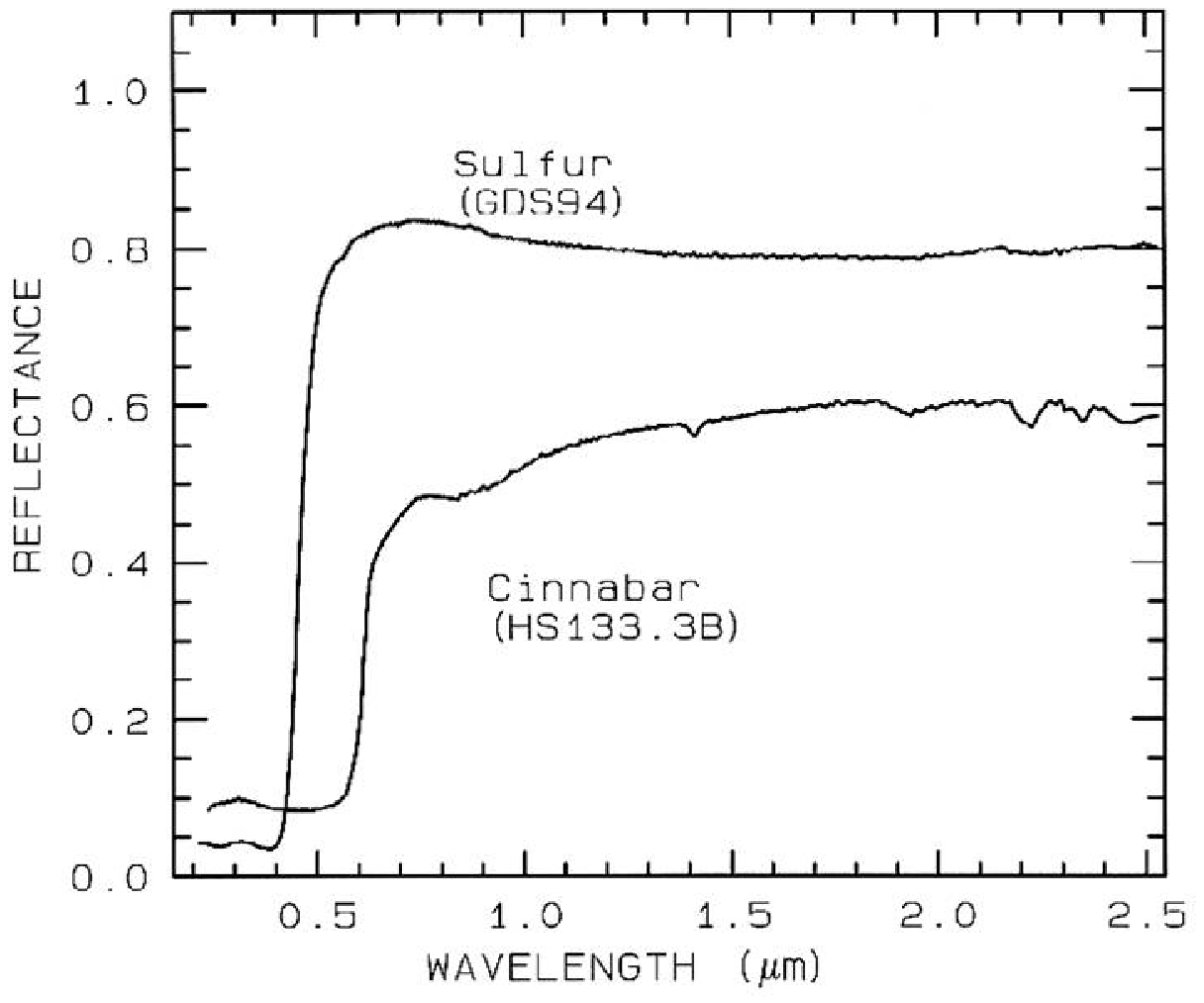}
\includegraphics[height=5cm]{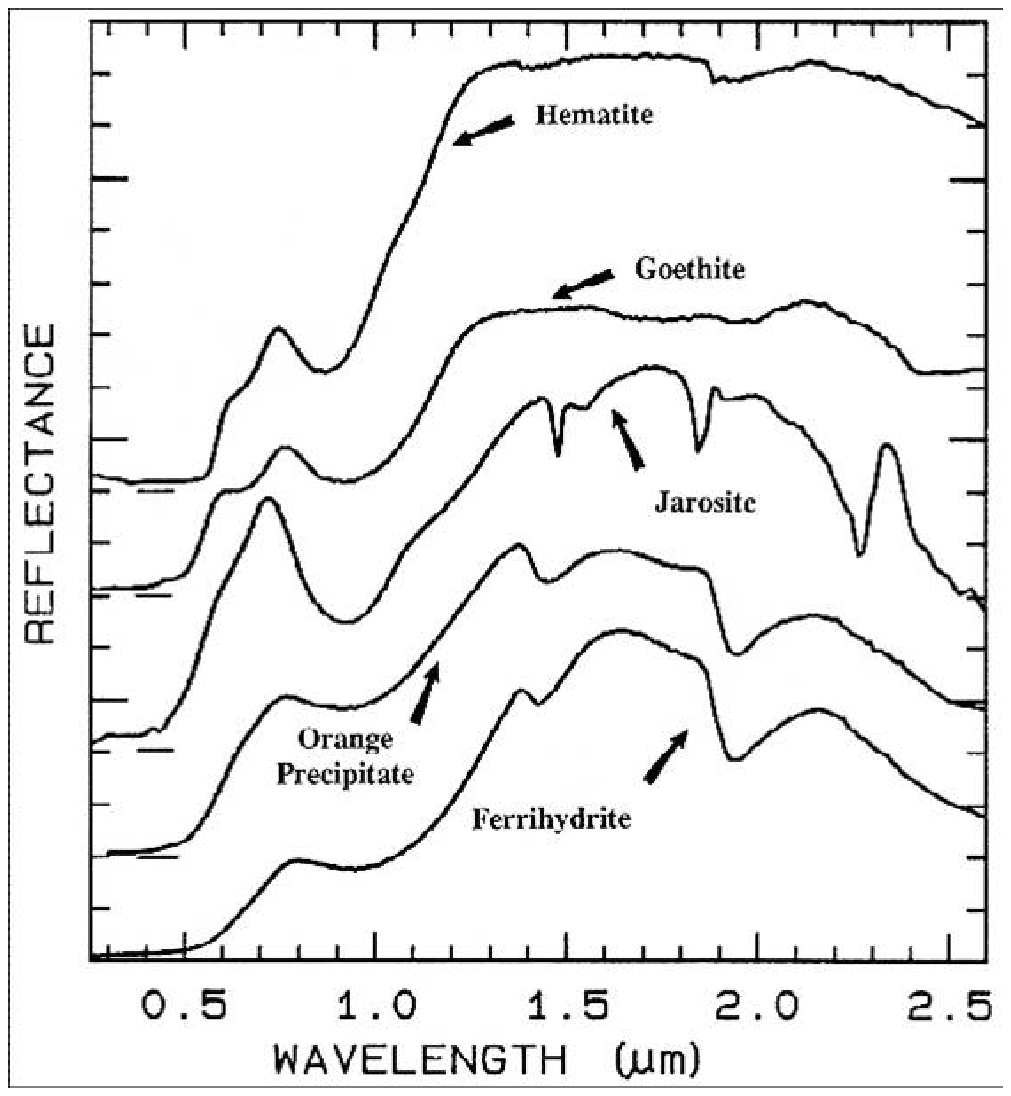}
      \caption{Minerals can display 'edges' in their reflectance spectrum. The possible confusion of the VRE with mineral spectral features has been discussed by Schneider (2004) and Seager et al. (2005), but no exhaustive work on this subject has been done to date. It would indeed be very interesting to know if the reflectance spectrum of the vegetation can be fitted by a relevant combination of spectra of minerals. Spectra are from \cite{clark99}.}
       \label{artifact}
\end{figure}

\textit{Is the red edge universal, i.e. inherent to any photosynthetic process in the Universe?}
Although on Earth most of photosynthetic species show a red edge around 700 nm resulting in a signature visible at a global scale, there are exceptions like \textit{Rhodopseudomonas} purple bacteria \cite{blankenship_et_al95}. Thus, strictly speaking, the red edge is not inherent to all photosynthetic species and thus probably not universal. On a life-detection strategy point of view, this observation suggests that, rather than looking for a earth-like red edge, we should look for a particular ground colour that could not be attributed to a mineral or a combination of minerals. If all mineral artifacts are eliminated, then only a photosynthetic process could be considered to interpret the spectrum (Fig.~\ref{artifact}).  

Considering that $O_2$ and $O_3$ are produced by photosynthesis on Earth, it seems thus relevant, if $O_2$, $O_3$ and $H_2O$ are detected in the spectrum of an Earth-like planet, to look for the signature of an extrasolar photosynthesis, i.e. a spectral feature - probably weak but hopefully sharp enough to be detectable and distinguishable from any known mineral- revealing missing photons used in a photosynthetic process. It seems relevant too to look for these missing photons at the wavelength were photons from the mother star are the most abundant, and also where the planetary atmosphere is the most transparent, so these photons can reach the ground. On the Earth, the atmosphere is indeed transparent to visible light and plants pigments involved in the photosynthesis strongly absorb in that spectral window. In order to access to the ground spectral signature, the atmosphere must be partially clear to allow us to \textit{see} the ground. It will be necessary to remove, at least partially, the atmospheric spectral bands to see that ground, meaning that we will need, at some stage, a model of the planet atmosphere. 

\section{Conclusion}

Earthshine observations have shown that Earth's vegetation is detectable in the Earth integrated spectrum. The vegetation signal is only a few percents (0 to 5\% range) above the continuum. One reason because the signal is weak is simply because vegetated areas are often covered by clouds. We speculated about the possible high VRE of a dry planet, i.e. with low cloud cover, but pointed out also that the VRE at 700 nm may not be a universal signature of plants on an extrasolar Earth-like planet.

\begin{figure}
\centering
\includegraphics{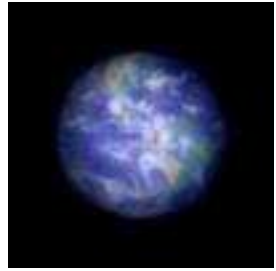}
\caption{Simulated image of the Earth at 10 light-years observed with a 150-km hypertelescope interferometric array made of 150 3-m mirrors working at visible wavelengths \cite{labeyrie99}. North and South America are visible. Note that this simulation is done at visible wavelengths, while in the (very) near-IR at 750 nm, vegetated areas would be much brighter and more easily detectable on continents. Spatial resolution at 750 nm would remain the same than at visible wavelength with the same hypertelescope flotilla being spread over 225-km instead of 150-km. }
\label{labeyrie99}
\end{figure}

Clearly \textit{resolved} images of extrasolar planets will help to detect photosynthetic life on these Earth-like planets!  But the wonderful instruments that will allow us to see Earth-like planets as small resolved disks are not yet ready to be launched (far from that), although possible designs have already been outlined. For example, a 150-km hypertelescope in space - an interferometric sparse array of small telescopes - would provide 40 resolution elements (resels) across an Earth at 10 light-years in yellow light \cite{labeyrie99}. And a formation of 150 3-m mirrors would collect enough photons in 30-min to freeze the rotation of the planet and produce an image with at least $\approx 300$ resels, and up to thousands  depending on array geometry (Fig.~\ref{labeyrie99}). At this level of spatial resolution, it will be possible to identify clouds, oceans and continents, either barren or perhaps (hopefully) conquered by vegetation.

\begin{acknowledgements}
The author acknowledges J. Schneider, Paris Observatory, S. Jacquemoud, Paris-7 University, for the stimulating discussions we had during the
writing of this paper, and the anonymous reviewer for the remarks that helped to improve the paper.
\end{acknowledgements}



\end{document}